\documentclass[useAMS,usenatbib]{mn2e}

\usepackage{graphicx,color}
\usepackage{aas_macros}

\title[Simultaneous polarimetric observations of SiO maser emission towards VY CMa]{Simultaneous VLBA polarimetric observations of the v=$\{$1,2$\}$ J=1-0 and v=1, J=2-1 SiO maser emission toward VY CMa: maser morphology and pumping}

\author[L. Richter et al.]
{
L.~Richter,$^1$\thanks{Present address: SKA South Africa, 3rd Floor, The Park, Park Road, Cape Town, 7405, South Africa}
A.~Kemball,$^{2,1}$\thanks{Visiting Professor affiliation} and
J.~Jonas$^1$
\\
  $^1$Department of Physics and Electronics, Rhodes University, Drostdy Road, Grahamstown, 6139, South Africa\\
  $^2$Department of Astronomy and Institute for Advanced Computing Applications and Technologies, University of Illinois at\\ Urbana-Champaign, 1002 W. Green Street, Urbana, IL, 61801, USA
}

\date{Released 2013 Xxxxx XX}

\pagerange{\pageref{firstpage}--\pageref{lastpage}} \pubyear{2013}

\def\LaTeX{L\kern-.36em\raise.3ex\hbox{a}\kern-.15em
    T\kern-.1667em\lower.7ex\hbox{E}\kern-.125emX}

\begin{document}

\label{firstpage}

\maketitle


\begin{abstract}
This paper presents a milliarcsecond-scale comparison of the polarised component-level v=1 J=1-0,
 v=2 J=1-0 and v=1 J=2-1 SiO maser emission toward the supergiant star VY CMa. These observations 
used the VLBA at $\lambda = 7$ mm and $\lambda = 3$ mm over two epochs. The goal is to use the 
relative characteristics and spatial distribution of the transitions in individual resolved maser 
components to provide observational constraints on SiO maser excitation and pumping models. We 
find that many v=1 J=1-0 and v=2 J=1-0  features overlap in the second observing epoch, at 
comparable sensitivity. The v=2 J=1-0 emission is primarily located within several v=1 J=1-0 
regions, and has a high degree of morphological similarity in the northeastern envelope, supportive 
of local collisional pumping. We find significantly higher spatial overlap between v=1 J=1-0 and 
J=2-1 features than generally previously reported. However, the overlapping v=1 J=2-1 emission 
is usually weaker than predicted by hydrodynamical models, possibly due to low-density collisional 
pumping.  The overall maser morphology contains several large-scale features that are persistent 
over multiple years and are likely associated with intrinsic physical circumstellar conditions. 
Our data cannot distinguish between competing near-circumstellar bulk kinematic models but do provide 
evidence for localized inhomogeneous mass-loss.
\end{abstract}



\section{Introduction}

SiO maser emission is commonly found in the atmospheres of late-type evolved stars \citep{Habing:96}, and has 
been used as a probe of their near-circumstellar envelopes. High resolution VLBI images of the masers provide 
information about circumstellar envelope kinematics \citep{Boboltz:97,Diamond:03,Zhang:12}, and inform
hydrodynamical models of the region \citep{Humphreys:96,Humphreys:02b,Gray:09}. 

However, the predominant pumping mechanism maintaining the circumstellar masers is still a subject of debate. 
Both radiative and collisional pumping mechanisms have been proposed \citep{Kwan:74,Elitzur:80}, and SiO maser 
hydrodynamical models have incorporated both, at varying levels of influence 
\citep{Humphreys:96,Humphreys:02b,Gray:09}. The SiO maser pumping models make predictions about the extent 
of spatial overlap between maser emission from different SiO transitions, so observational tests are important in 
informing theoretical work on SiO maser pumping.

This paper reports two epochs of full-polarisation Very Large Baseline Array (VLBA\footnote{The VLBA is operated by 
the National Radio Astronomy Observatory (NRAO). The NRAO is a facility of the National Science Foundation operated 
under cooperative agreement by Associated Universities, Inc.}) observations of SiO maser emission in the v=$\{$1,2$\}$ 
J=1-0 and v=1, J=2-1 transitions toward the highly luminous evolved star VY CMa. This source was chosen as the 
object of this study in spite of its complex circumstellar environment \citep{Smith:01}, but rather because it is such 
a strong SiO maser source, displaying emission from a wide range of SiO transitions \citep{Cernicharo:93}. Simultaneous 
observations of a range of SiO maser transitions in full polarisation at the component level provides stringent 
observational tests of SiO maser pumping models, and we present such an analysis here. The high SiO luminosity 
of VY CMa is also an important factor, given the sensitivity limitations of VLBI arrays at millimetre wavelengths.

Relevant predictions of the pumping models are discussed below, followed by a summary of existing 
comparative observations.

\subsection{SiO maser pumping models}

For both radiatively- and collisionally-pumped masers, the v=1 and v=2 J=1-0 transitions are inverted under 
different conditions \citep{Lockett:92,Bujarrabal:94a}. However, while the basic theory of radiative pumping 
predicts almost no spatial overlap between v=1 and v=2 transitions, they are predicted to show a significant 
amount of overlap for collisional pumping \citep{Lockett:92}. 
Considerable spatial overlap between the v=1 and v=2 \mbox{J=1-0} emission therefore 
argues for a collisional pumping mechanism. In the collisional pumping model 
of \citet{Lockett:92}, the range of conditions under which the v=1 J=1-0 masing occurs is wider than that of 
the v=2 J=1-0 transition, with the v=2 line requiring higher-density conditions.

Alternatively, extensive overlap between the v=1 and v=2 J=1-0 emission may also be caused by a line overlap 
between the v$_2$=0 $12_{75}$ $\to$ v$_2$=1 $11_{66}$ H$_2$O transition at 1219.10~cm$^{-1}$ and the adjacent 
v=1 J=0 $\to$ v=2 J=1 SiO transition at 1219.15~cm$^{-1}$ \citep{Olofsson:81}. This H$_2$O line 
overlap mechanism was originally invoked to explain the anomalously low v=2 J=2-1 SiO maser emission
generally found \citep{Olofsson:81,Bujarrabal:96}. \citet{Soria-Ruiz:04} investigated the effect of the proposed 
line overlap using the primarily radiative SiO maser excitation model of \citet{Bujarrabal:96}. A key result of 
their work is that when the line overlap is included in radiative transfer calculations, the conditions for v=1 and 
v=2 J=1-0 masers to occur intersect considerably, but the conditions under which the v=1 J=1-0 and and J=2-1 
masers would occur simultaneously are, in contrast, far more limited.

SiO maser excitation models have been coupled to hydrodynamical models of the circumstellar environment, 
and used to create simulated images of SiO maser emission in circumstellar envelopes 
\citep{Humphreys:96,Gray:00,Humphreys:02b,Gray:09}.
The models have had various successes, including predicting the existence of SiO maser emission from high 
angular momentum states above J=6 (\citealt*{Gray:95b},\citealt{Gray:95}), and predicting the ring morphology of the maser 
emission \citep{Humphreys:96}. Although these models do include a radiative pumping component, the pumping 
mechanism is predicted to be and treated as predominantly collisional \citep{Humphreys:96}. 

The hydrodynamical SiO maser models make a number of specific predictions about the relative location of 
SiO maser emission from the v=1 J=1-0, v=2 J=1-0 and v=1 J=2-1 transitions, and improve the predictive power 
of the multi-transition observational tests considered in this paper. 
The hydrodynamical model predictions are outlined 
in \citet{Humphreys:96}, \citet{Humphreys:02b} and \citet{Gray:09}, and include the following specific predictions:
\begin{itemize}
 \item The v=1 J=1-0 maser ring is thicker than that of the other two transitions, and maser emission occurs 
       over a wider range of physical conditions \citep{Gray:00}.
 \item The v=2 masers lie closer to the star than the v=1 masers \citep{Gray:00,Gray:09}. 
 \item The v=1 J=1-0 and J=2-1 lines often arise in shared components, in rings of similar radii, and the radial
       motions of the rings from these two lines are coupled \citep{Gray:00,Humphreys:02b}.
 \item Where v=1 J=2-1 and v=1 J=1-0 features overlap, the v=1 J=2-1 emission will be brighter \citep{Humphreys:02b}.
\end{itemize}

Unlike the comparative v=1 and v=2 J=1-0 case, the v=1 J=1-0 and J=2-1 SiO maser emission is expected to be
coincident from first principles under both radiative and collisional pumping schemes, as maser emission occurs along 
rotational ladders within a given vibrational state \citep{Alcolea:04}.
Lack of coincidence between v=1 J=1-0 and J=2-1 emission could be accounted for by a line overlap, as mentioned above,
or by variable envelope conditions favouring one or the other transition in a particular location.


\begin{table*}
\begin{minipage}{168mm}
\caption{Summary of the VLBA observations performed during epoch 1 (2003) and epoch 2 (2007).}
\begin{tabular}{|cccccccc|}
\hline 
Epoch & Transition & Frequency Band  & Clean Beam & Antennas & Stokes $I$ peak  & $\sigma_I^{B\,a}$ & Time on source \\
 & & [GHz] & [$\mu$arcsec] & & [Jy/beam] & [Jy/beam] & [min] \\

1 & v=1 J=1-0 & 43~GHz & $820\times170^b$ & 10$^d$ & 19.18 & 0.389 & 41 \\
1 & v=2 J=1-0 & 43~GHz & $790\times170^b$ & 10$^d$ & 45.75 & 0.091 & 41 \\ 
1 & v=1 J=2-1 & 86~GHz & $1220\times360^b$ & 8$^e$ & 59.93 & 0.702 & 23 \\ 
2 & v=1 J=1-0 & 43~GHz & $460\times150^c$ & 10$^f$ & 22.15 & 0.089 & 150 \\
2 & v=2 J=1-0 & 43~GHz & $430\times140^c$ & 9$^g$ & 12.16 & 0.134 & 150 \\ 
2 & v=1 J=2-1 & 86~GHz & $420\times90^c$ & 8$^h$ & 46.96 & 0.200 & 150 \\ 
\hline
\end{tabular}
%
$^a$Broadened estimate of the rms thermal noise per spectral channel, as described in Section~\ref{sec:Observations};
$^b$CLEAN beam in natural gridding weighting;
$^c$CLEAN beam in uniform gridding weighting;
$^d$VLBA antennas: BR, FD, HN, KP, LA, MK, NL, OV, PT, SC;
$^e$VLBA antennas: FD, HN, KP, LA, MK, NL, OV, PT;
$^f$VLBA antennas: BR, HN, KP, LA, MK, NL, OV, PT, SC, and a single VLA antenna;
$^g$VLBA antennas: BR, HN, KP, LA, MK, NL, OV, PT, SC;
$^h$VLBA antennas: BR, FD, KP, LA, MK, NL, OV, PT.
\label{table:observation_summary}
\end{minipage}
\end{table*}

\subsection{Previous comparative observations}

Simultaneous imaging VLBI observations of the v=1 and v=2 J=1-0 masers have previously been performed towards numerous 
late-type evolved stars. The earliest maps, presented by \citet{Miyoshi:94} in total intensity, show overlap between 
maser features in these two lines. A subsequent higher-resolution total intensity and linear polarisation observation 
by \citet{Desmurs:00} showed a systematic offset between emission from the two transitions, which they put forward as 
evidence in support of radiative pumping models.

A number of subsequent observations at similar angular resolution to those of \citet{Desmurs:00} showed significant 
overlap between the v=1 and v=2 J=1-0 masers \citep{Miyoshi:03,Yi:05,Cotton:06,Cotton:08}. In the overlapping v=1 
and v=2 features present in the \citet{Yi:05} maps, the v=2 emission tends to arise closer to the star than the v=1 
emission, with an intermediate region of overlap. 

More recent astrometrically-aligned maps of the v=1 and v=2 J=1-0 emission towards R Aquarii produced by the 
VERA array argue that the number of coincident maser spots in these two lines is actually small, but that v=1 and 
v=2 spots are clustered together and may appear coincident at lower resolution, or if absolute position information 
about the images is not available \citep{Kamohara:10}. This supports the \citet{Desmurs:00} result. \citet{Kamohara:10} 
find a number of spot clusters that have offsets of 1-2~mas between the v=1 and v=2 J=1-0 emission. However, the 
sensitivity of these VERA images is considerably lower than the VLBA images presented in several of the publications 
discussed above, so it is possible that weaker overlapping emission may not have been detected in the \citet{Kamohara:10} 
observations.

Simultaneous observations of the v=1 J=1-0 and v=1 J=2-1 SiO masers are less common than those of v=1 and v=2 
J=1-0 SiO masers, due to the limited number of 86~GHz observing facilities and the technical challenges posed by 
observing at these frequencies. Most of the published simultaneous VLBI imaging of these lines to date display no clear evidence 
of overlapping maser emission \citep{Desmurs:02b,Soria-Ruiz:04,Soria-Ruiz:05a,Soria-Ruiz:06,Soria-Ruiz:07}.
R~Cassiopeiae is an exception, and displays overlapping v=1 J=1-0 and v=1 J=2-1 SiO maser features in a total-intensity 
VLBI map published by \citet{Phillips:03}.

\subsection{Outline of this work}

We present a component-level comparison of the total intensity and linear polarisation properties of the SiO maser emission 
in the transitions v=$\{$1,2$\}$ J=1-0 and v=1, J=2-1 toward VY CMa, as well as an analysis of the overall morphology of 
the SiO maser emission toward this source.

We find that the v=2, J=1-0 emission is predominantly located within a subset of the v=1, J=1-0 emission 
regions overall, consistent both with the hydrodynamical models of \citet{Gray:00} as well as a predominantly 
collisional pumping interpretation. In addition, we find a higher degree of overlap between v=1 J=1-0 and 
v=1 \mbox{J=2-1} maser emission components than reported in the literature for other sources in the past. This 
is consistent with a model by \citet{Doel:95} and first-principle expectations for excitation of these transitions. 
The overall SiO morphology shows persistent features over multi-year intervals, that are very likely associated 
with intrinsic physical properties of the near circumstellar environment. We also find significant evidence for 
inhomogeneous and asymmetric localised mass-loss in this environment.

The organisation of the paper is described in what follows.
The VLBA observations and their reduction are described in Section~\ref{sec:Observations} and the results are presented 
in Section~\ref{sec:Results}. The results are discussed in Section~\ref{sec:Discussion}, in terms of the total intensity
maser maps (Section~\ref{subsec:Discussion_Intensity}), SiO maser pumping (Section~\ref{subsec:Discussion_Pumping}), 
inter-comparisons of the v=1 J=1-0 and v=2 J=1-0 (Section~\ref{subsec:Discussion_v=1_v=2}) and v=1 J=1-0 and v=1 
J=2-1 (Section~\ref{subsec:Discussion_J=1-0_j=2-1}) transitions respectively, and the larger-scale morphology of the 
circumstellar envelope (Section~\ref{subsec:Discussion_envelope}). The conclusions are presented in 
Section~\ref{sec:Conclusions}.

\section{Observations and Data Reduction}
\label{sec:Observations}

Two epochs of VLBA observations were performed in this study, conducted on 20 and 23 December 2003 (project 
BK103) and 15 and 19 March 2007 (project BR123). These two observations sets, under project codes BK103 and 
BR123, will be be denoted as epoch 1 and epoch 2 respectively, in what follows. Epoch 1 included observations of 
transitions $^{28}$SiO v=$\{$0,1,2$\}$ J=1-0 and J=2-1, $^{29}$SiO v=$\{$0,1,2$\}$ J=1-0 and J=2-1, and 
$^{30}$SiO v=$\{$0,1$\}$ J=1-0 and J=2-1. Of these, only transitions $^{28}$SiO v=$\{$0,1,2$\}$ J=1-0 and J=2-1 
were successfully imaged \citep[the total-intensity $^{28}$SiO v=$\{$1,2$\}$ J=1-0 images were presented in preliminary 
work by][]{Richter:07}. The other transitions had significantly lower signal-to-noise ratio and could not be imaged at 
sufficient dynamic range and are accordingly not included in this analysis.

To maximise the amount of time observing each transition, only transitions $^{28}$SiO v=$\{$0,1,2$\}$ J=1-0 and J=2-1, 
$^{29}$SiO v=1 J=1-0 and $^{30}$SiO v=0 J=1-0 were subsequently included in the epoch 2 schedule. Of these transitions,
we consider only $^{28}$SiO v=$\{$1,2$\}$ J=1-0 and v=1 J=2-1 here. 

Images of v=1 J=1-0, v=2 J=1-0 and v=1 J=2-1 $^{28}$SiO transitions from both epochs are presented in this paper. 
These lines were observed at adopted rest frequencies of 43122.03, 42820.48 and 86243.37~GHz respectively \citep{Muller:05}. 
The spectral windows for each line transition were centred in frequency assuming a systemic LSR velocity of +18 km.s$^{-1}$ for the 
target source VY~CMa. 

In the first epoch, the continuum extragalactic sources J0359+5057 and 3C273 were used as bandpass and phase 
calibrators for the 86~GHz observations, while the sources J0423-0120 and 3C273 were used in this role for the
43~GHz observations. At the second epoch, the sources 3C454.3, J0423-0120, J0609-1542 and 3C273 were used 
as bandpass and phase calibrators for all frequency bands.
		 	 	 		
The data were sampled using two-bit quantisation, and correlated in full cross-polarisation.  In epoch 1 the 43 GHz lines 
were recorded in spectral windows of bandwidth 4 MHz each and those in the 86 GHz band were recorded in 8 MHz 
spectral windows, yielding a comparable velocity range for the transitions in each band ($\triangle\nu = 27.8$~km.s$^{-1}$).
Analysis of the first epoch observations showed that the SiO maser emission extended beyond this observed velocity 
range in all of the observed transitions. The epoch 2 observations were consequently observed using double the spectral  
line bandwidth in each band over that used in epoch 1 (8~MHz at 43~GHz and 16~MHz at 86~GHz). In the epoch 2 
observations the total velocity range in the v=$\{$1,2$\}$ J=1-0 transitions was determined to be approximately $\sim40$~km.s$^{-1}$ 
about the systemic velocity. At the time of these observations, dual cross-polarisation correlation carried a concomitant 
limitation of 128 frequency channels per spectral window, and both epochs were correlated within this constraint. The nominal 
channel width at epoch 1 and epoch 2 was accordingly $\sim0.22$ and $\sim0.43$~km.s$^{-1}$ respectively.


\begin{figure}
\includegraphics[width=84mm]{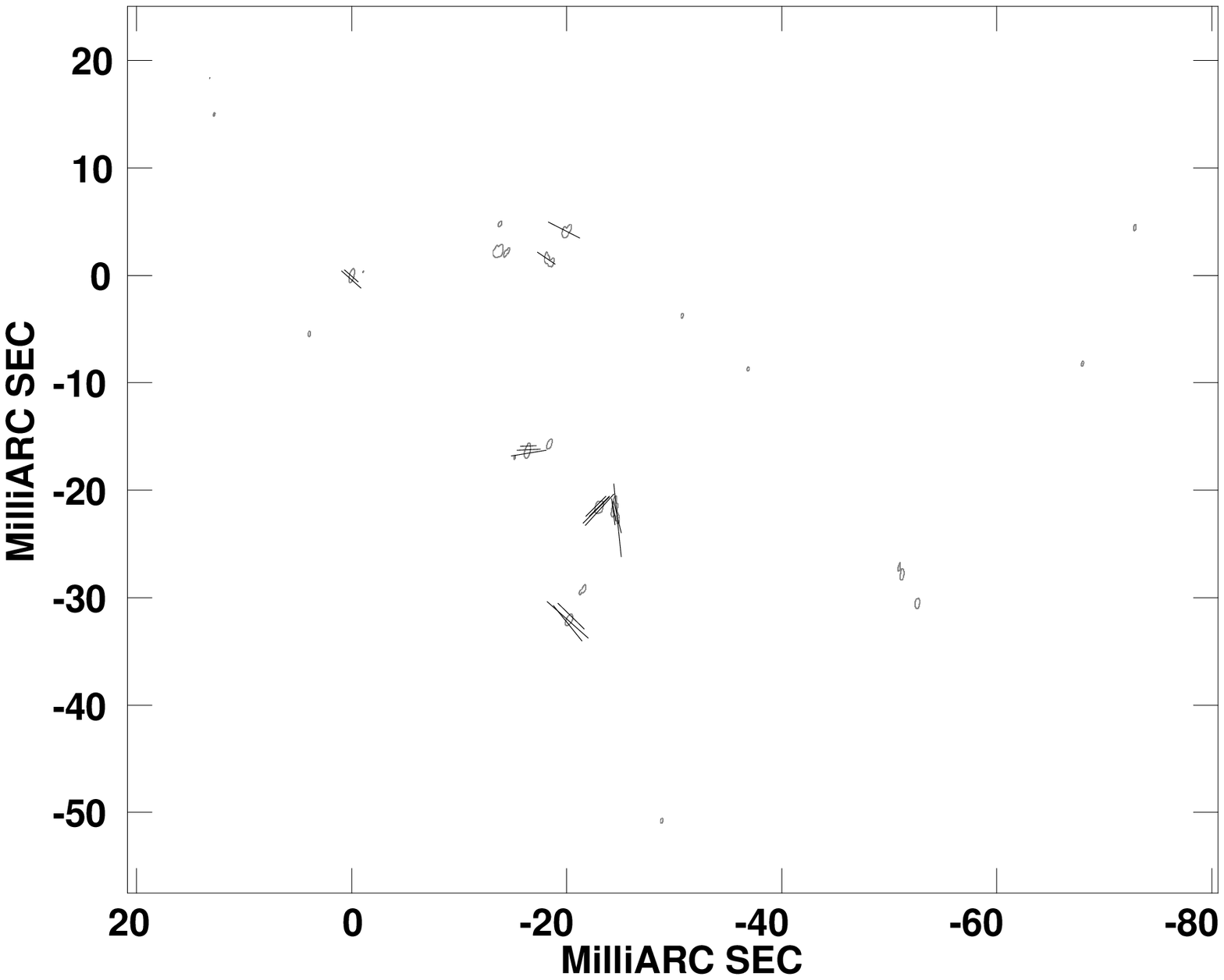}
\caption{Single-contour plot of the epoch 1 total intensity v=1 J=1-0 SiO maser emission toward VY CMa, as the peak intensity over the frequency 
axis of the cube, at a single level of $10\sigma_I^B$,  overlaid with vectors representing the averaged fractional linear polarised emission. 
The vectors are drawn with a length proportional to the magnitude of the fractional linear polarisation and an orientation aligned with the 
absolute EVPA of the underlying linearly polarised emission. The vector length scale is \mbox{1mas = $5.0\times10^{-2}$}.
The image is $4096\times4096$ pixels in size, at a pixel spacing of $30\mu$arcseconds.
The synthesised beam size is $0.82\times0.17$~mas in half-power at a position angle $-11.01^\circ$, N through E.}
\label{fig:7-2-3-P}
\end{figure}


\begin{figure}
\includegraphics[width=84mm]{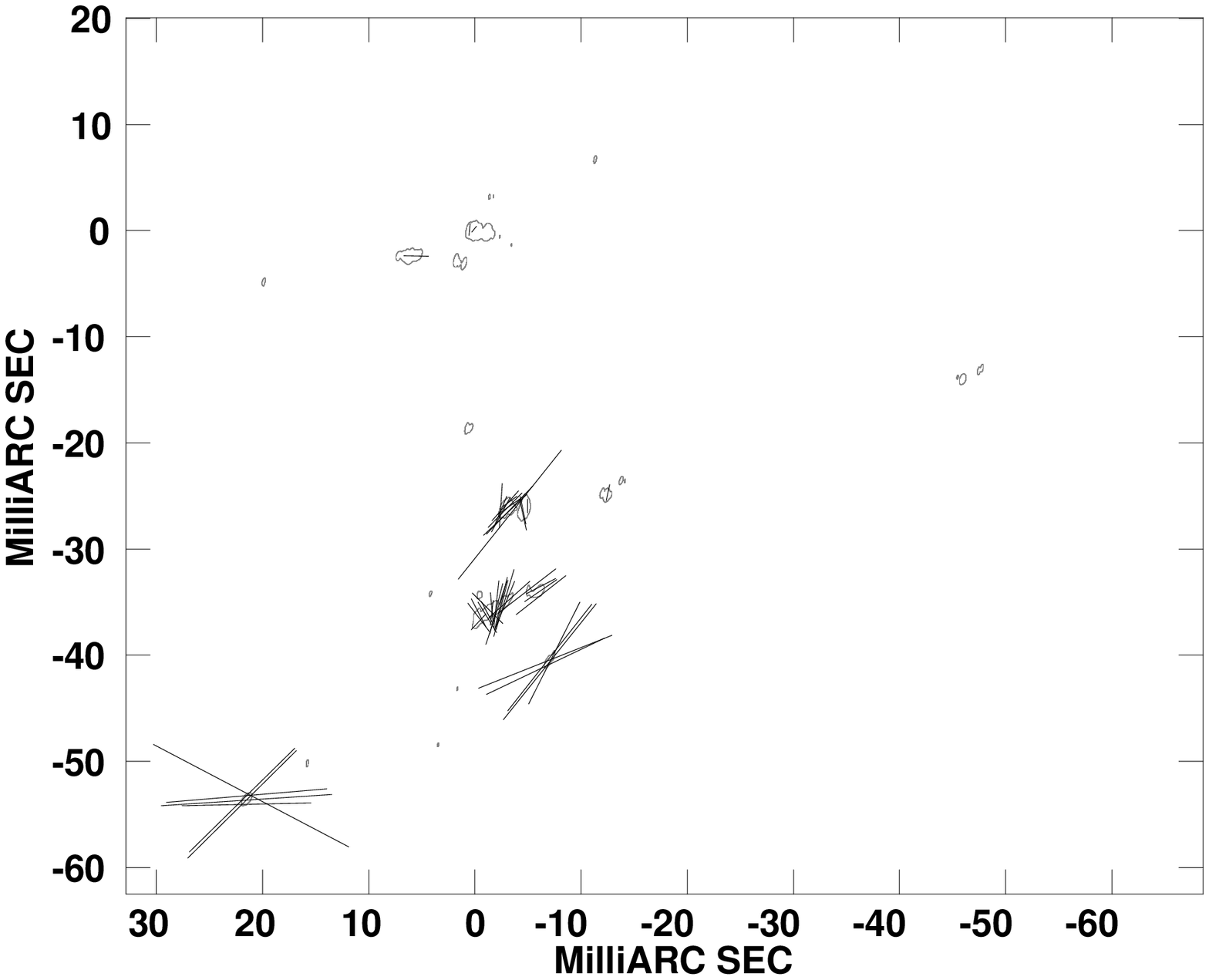}
\caption{Single-contour plot of the epoch 1 total intensity v=2 J=1-0 SiO maser emission toward VY CMa, as per 
Figure~\ref{fig:7-2-3-P}. The image parameters are the same as Figure~\ref{fig:7-2-3-P}, except for the synthesised beam size,
which is $0.79\times0.17$~mas in half-power at a position angle $-10.34^\circ$, N through E.}
\label{fig:7-2-1-P}
\end{figure}


\begin{figure}
\includegraphics[width=84mm]{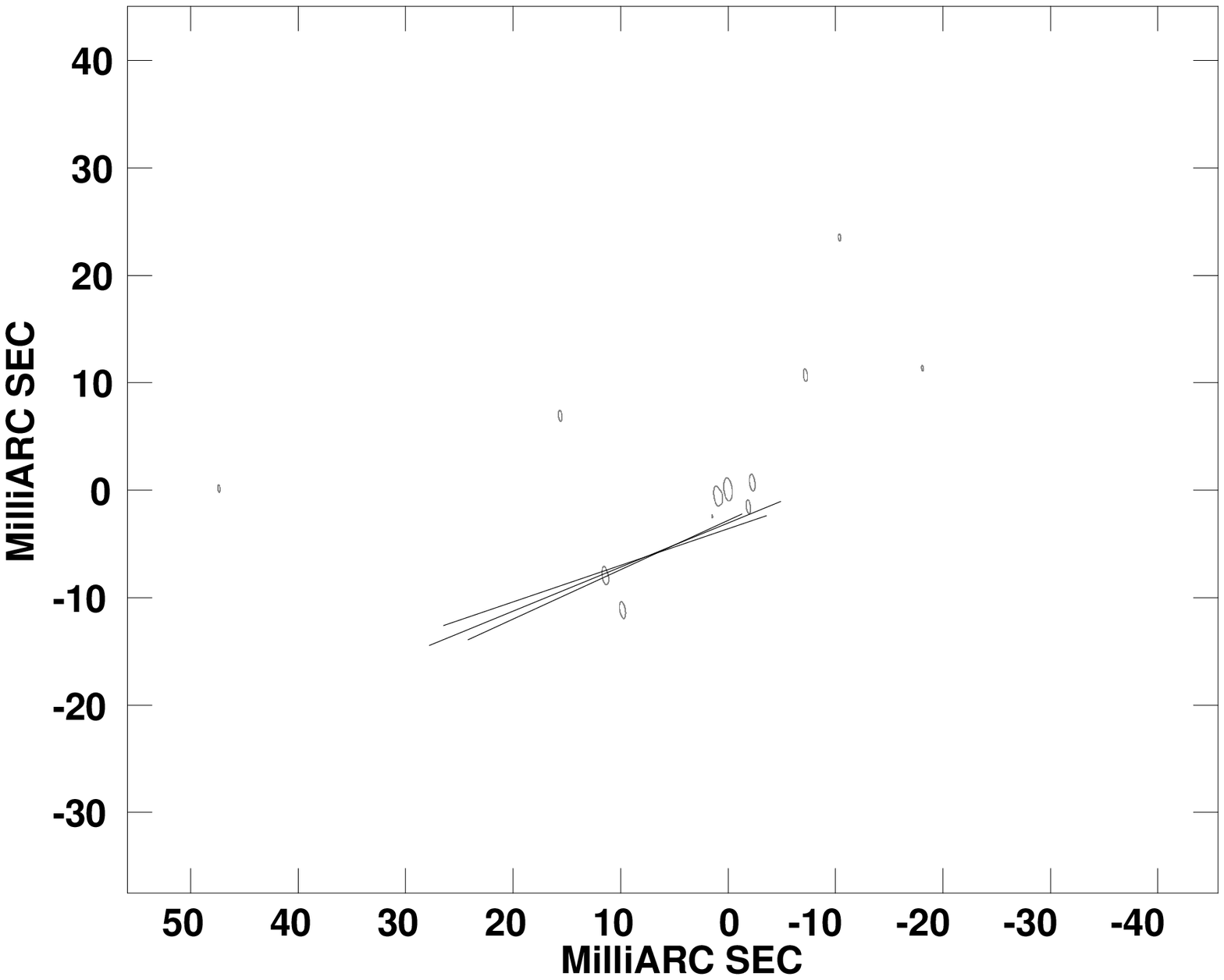}
\caption{Single-contour plot of the epoch 1 total intensity v=1 J=2-1 SiO maser emission toward VY CMa, as per
Figure~\ref{fig:7-2-3-P}. The image parameters are the same as Figure~\ref{fig:7-2-3-P}, except for the synthesised beam size,
which is $1.22\times0.36$~mas in half-power at a position angle $5.81^\circ$, N through E.}
\label{fig:3-5-1-P}
\end{figure}


\begin{figure}
\includegraphics[width=84mm]{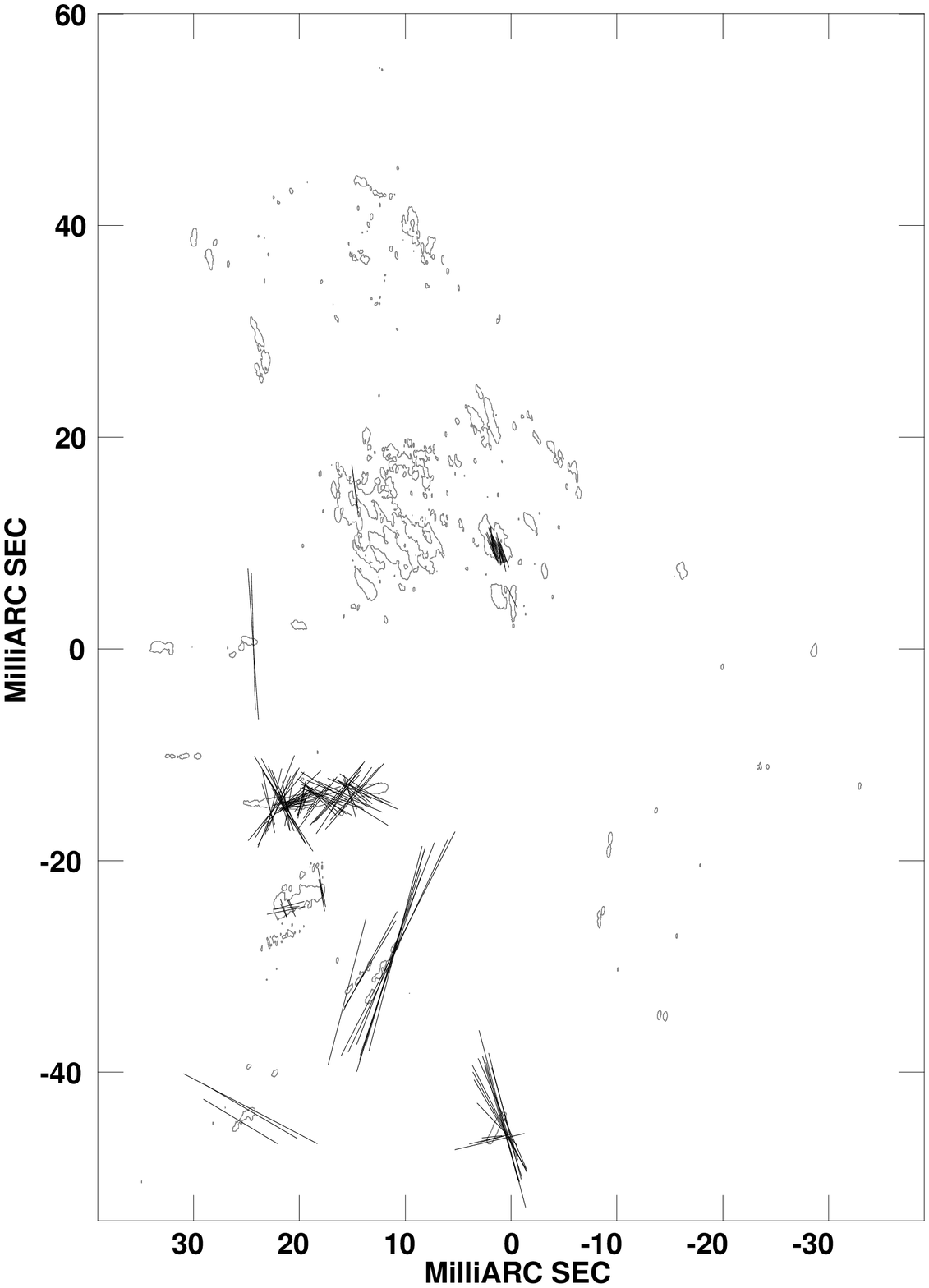}
\caption{Single-contour plot of the epoch 2 total intensity v=1 J=1-0 SiO maser emission toward VY CMa, as the peak intensity over the frequency 
axis of the cube, at a single level of $5\sigma_I^B$, overlaid with vectors representing the averaged fractional linear polarised emission. 
The vectors are drawn with a length proportional to the magnitude of the fractional linear polarisation and an orientation aligned with the 
absolute EVPA of the underlying linearly polarised emission. The vector length scale is \mbox{1mas = $2.5\times10^{-2}$}.
The image is $4096\times4096$ pixels in size, at a pixel spacing of $30\mu$arcseconds.
The synthesised beam size is $0.46\times0.15$~mas in half-power at a position angle $-1.80^\circ$, N through E.}
\label{fig:D2-P}
\end{figure}


\begin{figure}
\includegraphics[width=84mm]{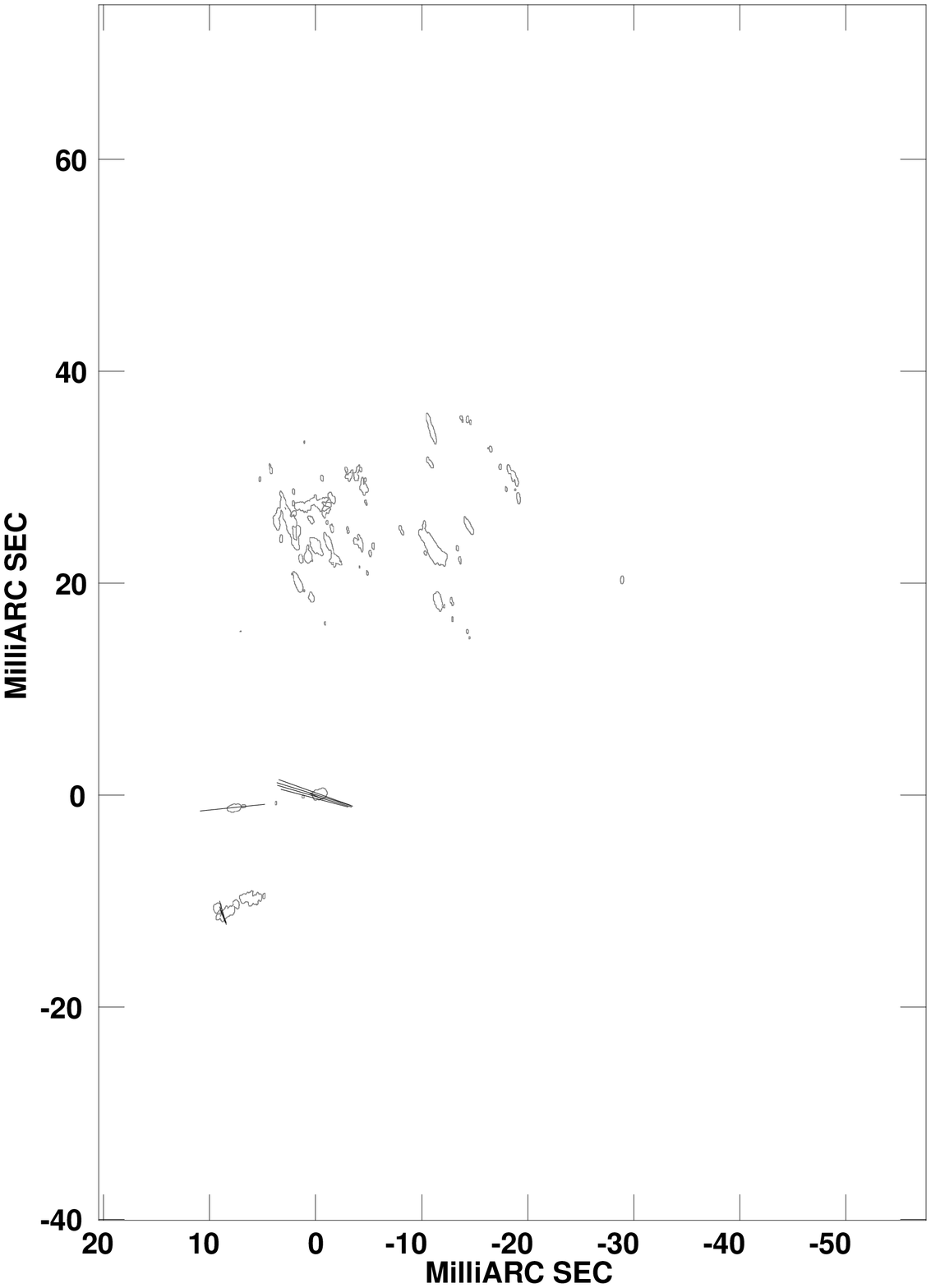}
\caption{Single-contour plot of the epoch 2 total intensity v=2 J=1-0 SiO maser emission toward VY CMa, as per 
Figure~\ref{fig:D2-P}. The image parameters are the same as Figure~\ref{fig:D2-P}, except for the synthesised beam size,
which is $0.43\times0.14$~mas in half-power at a position angle $-0.94^\circ$, N through E.}
\label{fig:D1-P}
\end{figure}


\begin{figure}
\includegraphics[width=84mm]{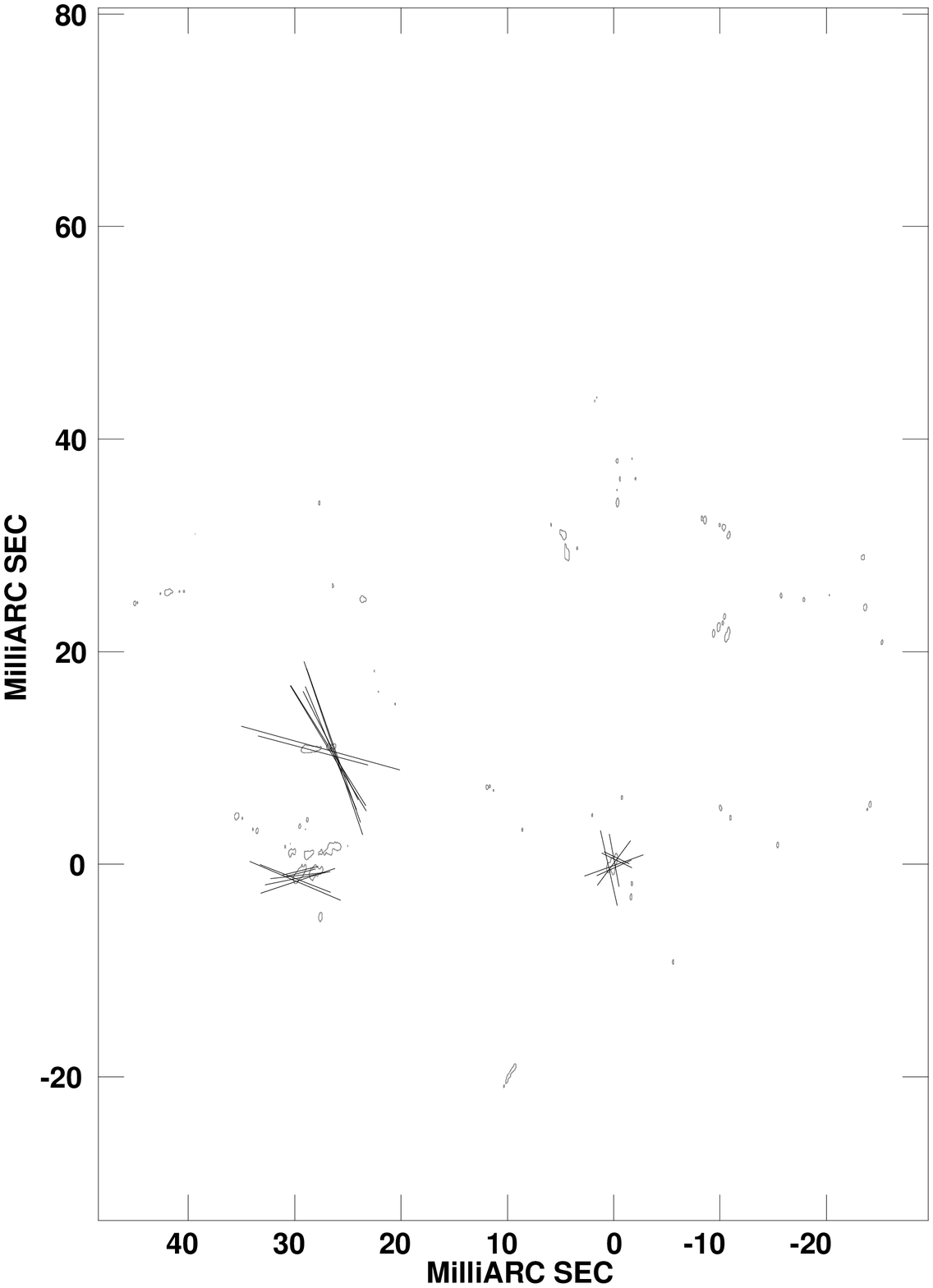}
\caption{Single-contour plot of the epoch 2 total intensity v=1 J=2-1 SiO maser emission toward VY CMa, as per
Figure~\ref{fig:D2-P}. The image parameters are the same as Figure~\ref{fig:D2-P}, except for the synthesised beam size,
which is $0.42\times0.09$~mas in half-power at a position angle $-16.35^\circ$, N through E.}
\label{fig:A-P}
\end{figure}


The epoch 1 data were reduced following the techniques outlined in \citet{Kemball:95} and \citet{Kemball:97}. These 
methods were refined further for high-precision Stokes $V$ measurement by \citet{Kemball:11}, and were then used to reduce the 
epoch 2 data in this work. For the total intensity and linear polarisation results reported in this paper, these two data 
reduction methods will not produce significantly different results. The data reduction was performed using a customised 
version of the Astronomical Image Processing System (\textsc{AIPS}\footnote{\textsc{AIPS} is developed and maintained by the NRAO 
(http://www.aips.nrao.edu)}). The observations are summarised in Table~\ref{table:observation_summary} for each transition 
at each epoch, listing the CLEAN restoring beam angular dimensions, the array configuration, the peak Stokes $I$ brightness 
(Jy/beam) in the resultant image cube, the broadened thermal noise estimate $\sigma_I^B$ (Jy/beam) (see below), and the total 
time on the target source VY CMa.
We note that the 86 GHz CLEAN beam sizes are larger than might be intuitively expected relative to the 43 GHz beam sizes; this is 
due to the absence of the outlying Saint Croix (SC) antenna in the 86 GHz observations.

The off-source RMS noise level in the maps is an under-estimate of the true noise in the maps, as residual calibration 
and deconvolution errors are direction-dependent and frequently more pronounced close to peaks in the emission. 
The noise estimates for the maps were therefore empirically broadened to a Gaussian $\sigma_I^B$ established from the
single-pixel deepest measured negative in the map, I$_{neg}$, and the number of pixels in the map, as described by 
\citet{Kemball:92}; in these channel images $\sigma_I^B=|\mbox{I}_{neg}|/5.295$.
The noise broadening is calculated per channel as the dynamic range varies with source structure over spectral channel.
The value of $\sigma_I^B$ quoted in Table~\ref{table:observation_summary}, and used to determine plotting
thresholds in subsequent sections, is the maximum value over all channels for each cube. The ratio of the broadened RMS noise 
to the measured RMS noise of the Stokes $I$ image, $\beta = \sigma_I / \sigma_I^B$, 
is also used to broaden the measured noise estimates of the Stokes $Q$ and $U$ maps: $\sigma_Q^B = \beta \sigma_Q$, 
$\sigma_U^B = \beta \sigma_U$ \citep{Kemball:92}. The unbroadened noise is similar in the Stokes $I$, $Q$ and $U$ maps, 
with broadening factors ranging between 1 and 11, with most typical values between 1 and 3. 
The broadened Stokes $Q$ and $U$ are conservative estimates of the Q and U noise however, due to the lower dynamic range 
limitations in linear polarisation images in general.

Values of the fractional linear polarisation $m_l$ quoted in subsequent sections were determined from the maximum
Stokes $I$ value of each feature, and the Stokes $Q$ and $U$ values at the corresponding pixel position of the Stokes 
$I$ measurement. The calculated fractional linear polarisation was corrected for Ricean bias, following \citet{Wardle:74}. 

The absolute electric vector position angle (EVPA) of the linear polarisation was determined through ancillary VLA
observations of a primary polarisation calibrator of known polarisation position angle \citep{Perley:03}, along with 
observations of compact secondary polarisation calibrators common to both the VLA and VLBA schedules. The VLA 
observations of the primary polarisation calibrators were used to determine the absolute EVPA of the secondary 
polarisation calibrators, which, by virtue of their inclusion in the VLBA schedule, could then be used to establish the 
absolute EVPA of the VLBA sources (or equivalently, the residual unknown R-L phase difference at the reference antenna, 
assumed constant) \citep{Kemball:99}.

The epoch 1 schedule included an auxiliary VLA observation on 20~December~2003, at which time the VLA was in 
B configuration. The VLA observed the primary polarisation calibrator J0521+166 (3C138) and secondary polarisation 
calibrator J0359+5057 in Q-band. The epoch 2 schedule included an auxiliary VLA observation on 17 March 2007. 
On this date the VLA was in D configuration. The calibrator 3C138 was again used as the primary polarisation calibrator 
in this epoch, and observed together with secondary polarisation calibrators J0646+448, J0609-1542, J0423-013 and 
J0542+498 in Q-band. It was not possible to perform absolute EVPA calibration of the 86~GHz data using this method, 
as the VLA is not equipped to observe at this frequency.

\section{Results}
\label{sec:Results}

Single-contour plots of the total intensity emission for each transition, as the peak over frequency across each image cube, 
and overlaid with vectors oriented at the EVPA and proportional in length to the underlying linearly polarised intensity, are 
shown for each data set in Figures~\ref{fig:7-2-3-P} to \ref{fig:A-P}. For the lower-sensitivity epoch 1 images, the single total 
intensity contour level was chosen as $10\sigma_I^B$. For the epoch 2 images, a contour level of $5\sigma_I^B$ was used.

Absolute astrometric information about the source is lost during the data reduction process, due to the use of phase 
self-calibration \citep{Thompson:04} 
(Equivalently, the origin of each of Figures~\ref{fig:7-2-3-P} to \ref{fig:A-P} is set by the choice of phase-reference channel 
and these processed maps are not intrinsically aligned).
The maser maps for each transition within each epoch were therefore aligned using a cross-correlation method, as described 
in Appendix~\ref{appendixA}.	 		
The uncertainty in the epoch 2 map alignment is estimated to be $< 0.05$ mas. The cross-correlation method required manual 
refinement for the epoch 1 maps, which had much fewer overlapping components. Assuming the correct peak has been selected, 
the uncertainty in the epoch 1 map alignment is also estimated to be $< 0.05$ mas.
In the alignment analysis, the epoch 2 deconvolved images were all restored with the same v=1 J=1-0 beam, whereas the
epoch 1 images were deconvolved with the intrinsic beam sizes listed in Figures~\ref{fig:7-2-3-P} to \ref{fig:3-5-1-P}; this 
is not believed to substantially affect the alignment method. 


\begin{figure}
\vspace{-0.7cm}
\includegraphics[width=90mm]{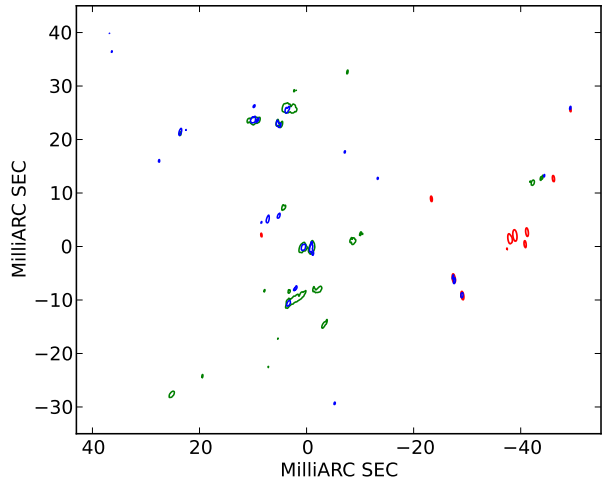} 
\caption{Overlaid epoch 1 total intensity contour plots of the SiO maser transitions v=1 J=1-0 (blue), v=2 J=1-0 (green), 
and v=1 J=2-1 (red) taken as the peak over frequency in each transition.
These images have been astrometrically aligned as described in the text.
A single contour is plotted for each transition, at a level of $10\sigma_I^B$ (see Table~\ref{table:observation_summary}).
}
\label{fig:I-overlay-epoch1}
\end{figure}

\begin{figure}
\includegraphics[width=84mm]{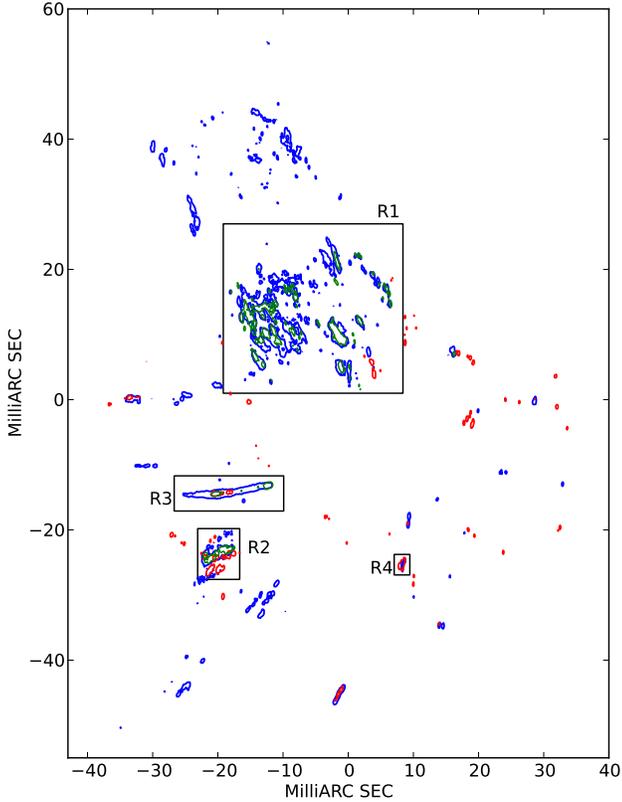}
\caption{Overlaid epoch 2 total intensity contour plots of the SiO maser transitions v=1 J=1-0 (blue), v=2 J=1-0 (green), 
and v=1 J=2-1 (red) taken as the peak over frequency in each transition.
These images have been astrometrically aligned as described in the text.
A single contour is plotted for each transition, at a level of $5\sigma_I^B$ (see Table~\ref{table:observation_summary}).
The regions R1, R2, R3 and R4 shown in the epoch 2 map are discussed in further detail in the text.
}
\label{fig:I-overlay-epoch2}
\end{figure}

The aligned total intensity maps are shown in Figures~\ref{fig:I-overlay-epoch1} and \ref{fig:I-overlay-epoch2}.
The epoch 2 v=1 J=1-0 emission is shown separately in Figure~\ref{fig:D2-velocity}, with contours colour-coded by velocity channel.
Enlarged plots of of several regions of the epoch 2 overlay map are shown in Figures~\ref{fig:I-overlay-R1-R2} and \ref{fig:I-overlay-R3-R4}. 

For the epoch 2 overlaid transition maps in Figures~\ref{fig:I-overlay-epoch2}, \ref{fig:I-overlay-R1-R2} and \ref{fig:I-overlay-R3-R4}, 
the deconvolved maps in each transition were restored with the same v=1 J=1-0 beam size. This aids their component-level interpretation (see 
Figures~\ref{fig:I-overlay-R1-R2} and \ref{fig:I-overlay-R3-R4} and associated discussion in the text). The single panel epoch 1 
overlay of multiple transitions (Figure~\ref{fig:I-overlay-epoch1}) uses the intrinsic beam sizes in Figures~\ref{fig:7-2-3-P} to 
\ref{fig:3-5-1-P}, as discussed above.


\section{Discussion}
\label{sec:Discussion}

\subsection{Intensity morphology}
\label{subsec:Discussion_Intensity}

For all of the total intensity maps presented in Figures~\ref{fig:7-2-3-P} to \ref{fig:A-P}, the emission falls within 
a region $\sim100\times100$~mas in angular extent. Adopting a stellar diameter of $18.7$~mas and a dust 
formation radius of $\sim40-50$~mas \citep{Danchi:94,Monnier:00}, most of the observed emission is located 
within a few stellar radii from the star, and within the dust formation radius.


\begin{figure}
\includegraphics[width=85mm]{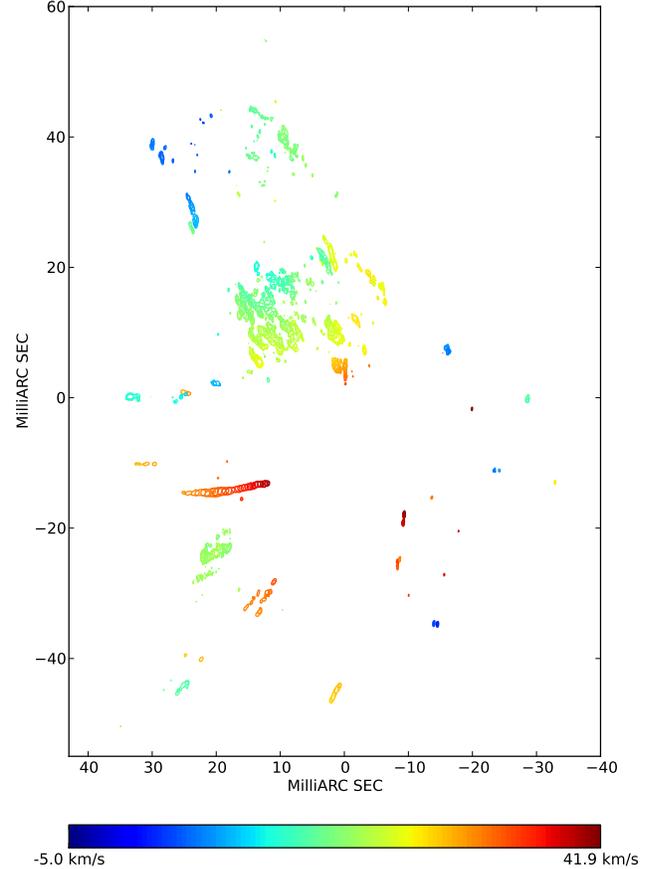} 
\caption{Epoch 2 total intensity contour plot of the SiO maser transition v=1, J=1-0, color-coded by line-of-sight 
velocity in the LSR frame as shown in the colour bar below the figure. The contour levels are -2, 2, 10, 20, 40, 60, and 
80$\%$ of the peak brightness of 22.15 Jy/beam.}
\label{fig:D2-velocity}
\end{figure}


Previous images of the SiO maser emission towards VY~CMa \citep{Miyoshi:94,Miyoshi:03,Shibata:04,Richter:07,Choi:08b,Zhang:12} 
show radially thicker and less well-defined rings than those often observed around AGB stars \citep{Diamond:94}.
The total intensity maps presented here follow the same trend: no clear ring structure is visible in the epoch~1 
maps, and the epoch~2 maps appear to show a sparse, wide ring. 


\begin{figure}
\includegraphics[width=84mm]{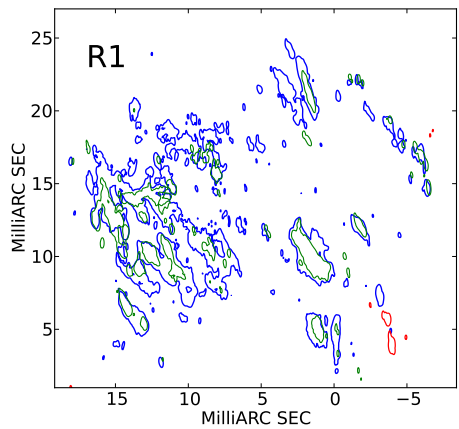}
\includegraphics[width=84mm]{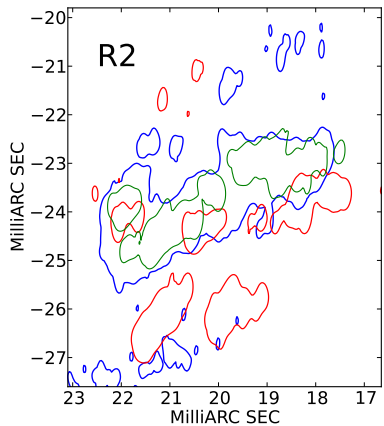}
\caption{As for Figure~\ref{fig:I-overlay-epoch2}, but showing the SiO maser transitions
v=1 J=1-0 (blue), v=2 J=1-0 (green), and v=1 J=2-1 (red) for epoch 2, for regions R1 and R2.}
\label{fig:I-overlay-R1-R2}
\end{figure}


\begin{figure}
\includegraphics[width=84mm]{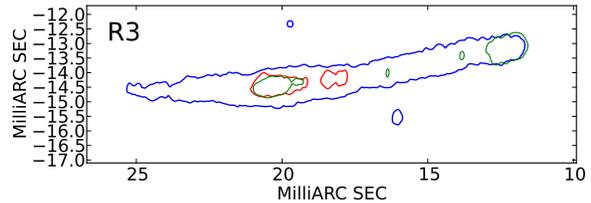}
\includegraphics[width=84mm]{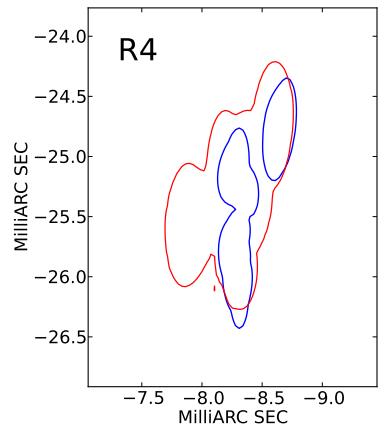}
\caption{As for Figure~\ref{fig:I-overlay-epoch2}, but showing the SiO maser transitions
v=1 J=1-0 (blue), v=2 J=1-0 (green), and v=1 J=2-1 (red) for epoch 2, for regions R3 and R4.}
\label{fig:I-overlay-R3-R4}
\end{figure}


The inner limit of the epoch 2 maser ring in the v=1 J=1-0 transition is not well defined, due to gaps in the emission.
The maximum radial extent is $\sim60$~mas in the northeast, with most of the emission around the rest of the ring 
falling within a radial extent of $\sim40$~mas from the approximate center of the emission. The \mbox{epoch 2} v=1 
J=1-0 maser distribution is very similar to images of the same transition 
presented in \citet{Zhang:12}, the most contemporaneous of which was observed about a month apart from our epoch 2. 
Multiple features can be visually matched between the epoch 2 maps presented 
here and the maps presented by both \citet{Choi:08b} and \citet{Zhang:12}. Both the v=1 and v=2 J=1-0 images 
display concentrated emission regions in the east and northeast (see regions R1-R4 in Figure~\ref{fig:I-overlay-epoch2}). 
One of the most striking features in the images is an $\sim14$~mas elongated feature in the eastern
part of the projected ring (R3 in Figure~\ref{fig:I-overlay-epoch2} and Figure~\ref{fig:I-overlay-R3-R4}), which is visible in 
maps presented by both \citet{Choi:08b} and \citet{Zhang:12}. These individual regions of emission are discussed 
in more detail in subsequent sections.

\begin{figure}
\hspace{-0.2cm}
\includegraphics[width=86mm]{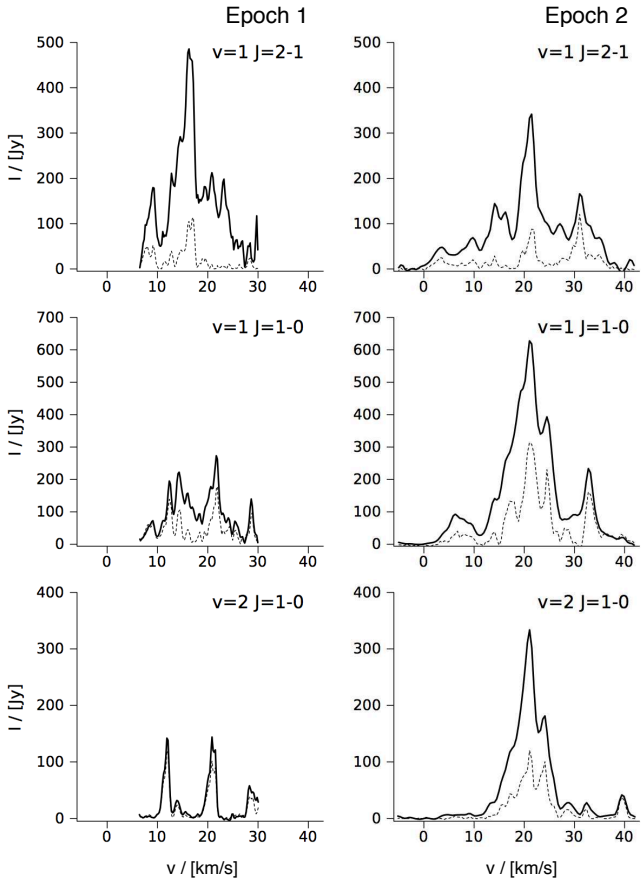} 
\caption{Single-dish total intensity autocorrelation spectra (solid line) and summed interferometric 
total intensity spectra (dashed line), for epoch 1 (left) and epoch 2 (right).
For epoch 1, the autocorrelation plots are high-SNR spectra from representative antennas (from top to bottom: Los Alamos, Los Alamos, Pie Town).
For epoch 2, the autocorrelation plots are composites of high-SNR spectra from multiple antennas.
}
\label{fig:AC-XC-spectra}
\end{figure}

The fraction of the total emitted maser flux recovered in the VLBA images can be determined through comparison of the 
interferometer autocorrelation spectra and the summed interferometric Stokes~$I$ spectra obtained from the final image cubes. 
These spectra are shown for each epoch in Figure~\ref{fig:AC-XC-spectra}.
In epoch 1, for the v=1 J=1-0 and v=2 J=1-0 lines $\sim50\%$ and 
$\sim75\%$ respectively of the integrated intensity from the autocorrelation spectra is recovered in the VLBA Stokes $I$ images.
The fraction recovered in the epoch 1 \mbox{v=1 J=2-1} line is much lower, at $\sim14\%$.
In epoch 2, approximately $35-40\%$ of the total emission from the \mbox{J=1-0} lines were recovered in the VLBA images.
For the epoch~2 v=1 J=2-1 line, approximately $22\%$ of the total emission was recovered interferometrically.

The J=1-0 fractions are consistent with VLBA observations of \citet{Cotton:06, Cotton:09a, Cotton:09b, Cotton:10a, Cotton:10b}
and \citet{Soria-Ruiz:04,Soria-Ruiz:05a,Soria-Ruiz:06,Soria-Ruiz:07}, who observed v=1 and v=2 \mbox{J=1-0} SiO maser emission 
towards a number of late-type stars, over multiple epochs in some cases.
These observations report flux recovery of a few tens of percent up to almost all of the single dish flux, varying across 
the spectra and over different epochs. 

The lower fraction of recovered flux for the J=2-1 line observed here is also consistent with previous VLBA observations 
of this transition, which show typical maximum flux recovery of $\sim10-15\%$ \citep{Soria-Ruiz:04,Soria-Ruiz:05a,Soria-Ruiz:07}.
A notable exception is \citet{Soria-Ruiz:06}, who recover up to $\sim50\%$ of the v=1 J=2-1 SiO maser emission towards
TX Cam, for individual features in the spectrum.

Missing flux can be attributed to diffuse emission, on a scale too large to be detected by the VLBA, or to numerous
low intensity maser spots which fall below the noise level of the images \citep{Gray:09}.
By definition, the spatial distribution of the missing flux is unconstrained by the sampled interferometer visibility data 
and therefore unknown \citep{Bracewell:54}.
The problem of missing flux impacts comparative studies at different frequencies, because the spatial scales recovered
may unavoidably differ per transition.
The differing brightness temperature sensitivity does not introduce a systematic spatial offset between transitions however, 
and therefore does not impair our key conclusions regarding the degree of spatial alignment and superimposition.

Emission at larger angular scales than the reciprocal of the shortest baseline (in wavelength-units) is not visible to 
interferometers \citep{Thompson:04}. The shortest unprojected VLBA baseline is 236~km, between antennas Los Alamos and Pie 
Town,\footnote{The epoch 2 v=1 J=1-0 observation did include one VLA antenna, providing a shorter baseline between Pie Town 
and the VLA antenna, but the VLA data were flagged out.} which corresponds to $\sim6.1$~mas at 43~GHz and 
$\sim2.6$~mas at 86~GHz. Emission on scales between $\sim3$ and 6~mas may therefore be detected by the 43~GHz observations 
presented here, but not by the 83~GHz observations.
For these data, the brightness temperature sensitivity at 43 GHz is approximately double that at 86 GHz.

The fact that more large-scale emission is filtered out in the 86 GHz observations is reflected in 
the recovered autocorrelation emission fraction, which is considerably lower at 86~GHz for both epochs. 
However, in the case of the extended northeastern region of emission in the epoch 2 J=1-0 maps, an argument can be made that
the absence of J=2-1 emission in the northwest is real. The epoch~2 autocorrelation and summed interferometric spectra show a 
spectrally-wide feature spanning about 15~km/s 
about the central velocity, which corresponds primarily to the more extended region of emission in the northeast of 
the J=1-0 epoch 2 maps. This is most clearly visible in Figure~\ref{fig:D2-velocity}, a contour plot of the v=1 J=1-0 maser 
emission colour-coded by velocity. The wide feature is not observed in the J=2-1 autocorrelation or summed interferometric
spectra.


\subsection{SiO maser pumping models}
\label{subsec:Discussion_Pumping}

As discussed earlier, models for circumstellar SiO maser excitation rely either on radiative pumping, collisional 
pumping, or a combination of the two \citep{Lockett:92,Bujarrabal:94a}. Which predominant pumping mechanism is 
driving circumstellar SiO masers remains an issue of some debate \citep{Desmurs:00,Soria-Ruiz:04,Gray:09}.
In this section we consider the relative characteristics of the v=1 J=1-0, v=2 J=1-0 and v=1 J=2-1 SiO maser 
emission, presented in Figures~\ref{fig:7-2-3-P} to \ref{fig:I-overlay-R3-R4}, as observational tests on the 
SiO maser pumping models discussed earlier.

\subsubsection{v=1 J=1-0 and v=2 J=1-0}
\label{subsec:Discussion_v=1_v=2}

Both epochs of VY~CMa observations presented here show overlap between the v=1 and v=2 J=1-0 emission. In 
the epoch~1 observations there are several v=2 features with no coincident, or even nearby, matching v=1 features.
However, during epoch~1 the noise level of the v=2 J=1-0 map was considerably lower than that of the v=1 J=1-0 map 
(Table~\ref{table:observation_summary}), so the more limited overlapping v=1 emission in epoch~1 is likely an SNR 
artifact.

In the epoch~2 observations, where the v=1 and v=2 J=1-0 maps have similar noise levels 
(Table~\ref{table:observation_summary}), the v=2 emission is located almost exclusively in a subset of v=1 emission 
regions. This is consistent with the \citet{Gray:00} hydrodynamical model, which predicts a thicker maser ring for the 
v=1 J=1-0 line, with the v=2 masers arising in a subset of the v=1 conditions.

In the extended regions of emission R1 and R2 (Figure~\ref{fig:I-overlay-epoch2}, Figure~\ref{fig:I-overlay-R1-R2}) the 
overlapping v=1 and v=2 J=1-0 maser emission is morphologically very similar. Most of the v=2 emission appears to 
follow the v=1 emission closely in these regions, occupying the same regions, or subsets of regions, as those 
exhibiting v=1 emission. The extensive overlap in R1 especially argues either for a predominantly collisional pumping 
mechanism at work in this region, or if the pumping mechanism is radiative, then the overlap may be caused by the 
H$_2$O line overlap mechanism coupling the v=1 and v=2 J=1-0 lines \citep{Soria-Ruiz:04}. However, the linear 
polarisation is weak in region R1. 
Statistically significant linear polarisation is only measured in four v=1 J=1-0 
features, and no v=2 J=1-0 features, in region R1 (the statistical significance is measured using the broadened 
noise in Stokes $I$, $Q$ and $U$ defined above.) The four features which do display linear polarisation in v=1 J=1-0 are all
less than $10\%$ polarised. 
The overall maximum linear polarisation across all transitions is $\sim45\%$ in epoch 2. The component-level polarisation 
characteristics will be presented in greater detail a future publication.

Linear polarisation is a natural characteristic of anisotropic radiative pumping from 
a central star \citep{Bujarrabal:81}. This effect would also tend to be weaker further from the star, as stellar radiation 
is diluted over a greater volume. So the wide radial extent of the maser emission in region R1 also argues against strong 
anisotropic radiative pumping in this region. 

The close coupling of the v=1 and v=2 J=1-0 maser emission in R1 may 
still be evidence of a H$_2$O line overlap impacting the pumping scheme, but possibly in the context of a predominantly 
collisional pumping mechanism. In R2, if the v=1 and v=2 J=1-0 emission overlap is primarily due to the H$_2$O line 
overlap mechanism then the observed spatial overlap between the v=1, J=1-0 and v=1, J=2-1 lines would only occur 
under a very restricted set of envelope conditions \citep{Soria-Ruiz:04}.

The component spectrum for the elongated feature R3 (Figure~\ref{fig:I-overlay-epoch2} and Figure~\ref{fig:I-overlay-R3-R4})
is shown in Figure~\ref{fig:R3}. 
The v=2 J=1-0 frequency axis in Figure~\ref{fig:R3} is shifted by two channels ($\sim0.86$~km.s$^{-1}$) to account for an 
observed $\sim2$ channel frequency offset between the positions of maser features in the v=1 and v=2 J=1-0 transitions. 
The offset is likely due to errors in the assumed rest frequencies of these transitions. The rest frequencies used in the 
data reduction were taken from ``The Cologne Database for Molecular Spectroscopy'' \citep{Muller:05}, whereas the images of 
overlapping components are more consistent with the frequencies catalogued in the ``Spectral Line Atlas of Interstellar 
Molecules'' \citep{Remijan:07}. High precision velocity information is not required by this work, so precise corrections 
for the offset were not attempted.

The v=2 J=1-0 emission in R3 is located within two main subsets of the v=1 J=1-0 emission region: near the region of peak 
intensity, and at the inner end of the feature, closest to the star. Where the v=1 and v=2 J=1-0 emission overlaps, it is 
expected that the v=2 emission will occur closest to the star, under a radiative pumping scheme \citep{Desmurs:00}, or in 
higher density conditions (also presumably closer to the star), under a collisional pumping scheme \citep{Doel:95,Gray:00}. 
The v=2 emission at the inner end of the long feature can therefore be explained by either pumping mechanism. The overlapping 
region in the peak intensity centre of the R3 may be due to the coherent path length of the maser emission being longest in 
this region, allowing significant amplification for all three maser transitions.
Alternatively, the overlapping region of emission in R3 may be a result of competitive gain, where transitions in different 
vibrationally-excited states become coupled as the maser emission saturates \citep{Doel:95}. 
The effect of saturation on maser polarisation varies between maser polarisation models, with anisotropic pumping models
showing linear polarisation decreasing with saturation \citep{Nedoluha:90}, some maser polarisation modelling showing
increasing linear polarisation with saturation \citep{Western:84} and other modelling showing that the level of linear
polarisation is the same in the unsaturated and saturated regimes, all other factors being equal \citep{Elitzur:91}.
We note here that feature R3 has statistically significant linear polarisation in all three transitions, at a level of
$\sim1$ to $20\%$ for the J=1-0 transitions, and $\sim25$ to $40\%$ for the J=2-1 transition. A detailed presentation of 
the component-level polarisation properties and their implications will be presented in a future publication.

\subsubsection{v=1 J=1-0 and v=1 J=2-1}
\label{subsec:Discussion_J=1-0_j=2-1}


\begin{figure}
\includegraphics[width=84mm]{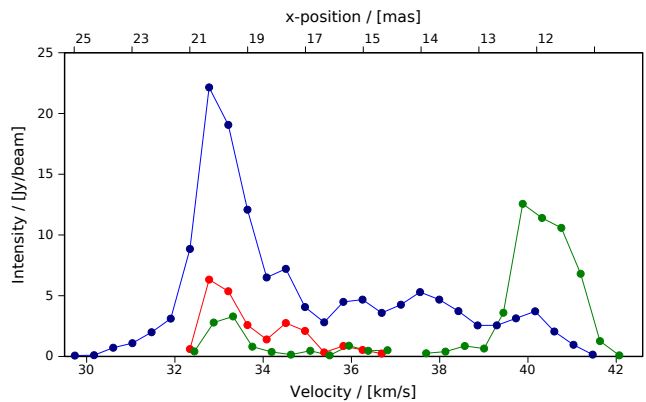}
\caption{Stokes $I$ spectra of feature R3 (Figure~\ref{fig:I-overlay-epoch2}), 
plotted against line of sight velocity in the Local Standard of Rest frame (bottom) and position along
the maser feature (top).
The colour mapping for each SiO maser transition follows that used in Figure \ref{fig:I-overlay-R3-R4}.}
\label{fig:R3}
\end{figure}


As noted earlier, simultaneous measurements of the spatial overlap of v=1, J=1-0 and v=1, J=2-1 SiO masers are 
limited in number due to technical challenges and show mixed results in terms of the degree of spatial overlap 
\citep{Desmurs:02b,Phillips:03,Soria-Ruiz:04,Soria-Ruiz:05a,Soria-Ruiz:06,Soria-Ruiz:07}. Generally little overlap 
has been found.

Unlike these earlier results, the maps shown in Figure~\ref{fig:I-overlay-epoch1} and \ref{fig:I-overlay-epoch2} 
show several overlapping maser features.
These are observed in both epochs, but in each case only a small subset of the total number of maser features overlap. 
We note however that the image noise levels in these two transitions are not equal (see Table~\ref{table:observation_summary}), 
a caveat which must be borne in mind in the discussion that follows.
Most of the v=1 J=1-0 features have no counterpart in v=1 J=2-1, and \textit{vice versa}. The large-scale northeastern 
region R1 of J=1-0 maser emission in the epoch~2 maps (Figure~\ref{fig:I-overlay-epoch2} and Figure~\ref{fig:I-overlay-R1-R2}) 
is also notably devoid of J=2-1 emission.

The spectra of the overlapping features show similar shape in the J=1-0 and J=2-1 lines, with the J=2-1 emission 
weaker in almost all cases. The similar spectral shapes provide evidence that the coincident maser emission from 
these two transitions very likely arises from the same physical regions of gas. One of the clearest examples is provided 
by feature R3 due to its large velocity extent; this component spectrum for all three transitions is shown in 
Figure~\ref{fig:R3}.

The coincidence of v=1 J=1-0 and J=2-1 SiO maser emission is expected from first principles, as noted above \citep{Alcolea:04}. 
However, the relative weakness of the \mbox{J=2-1} masers relative to the J=1-0 masers conflicts with the 
\citet{Humphreys:02b} model prediction of greater intensity in the J=2-1 line within overlapping components.

Under the predominantly collisional pumping SiO maser model of \citet{Doel:95}, the v=1 \mbox{J=2-1} maser 
amplification is greater than that of the v=1 \mbox{J=1-0} masers for intermediate densities around 
\mbox{$5\times10^9$~H$_2\;$cm$^{-1}$}. The v=1 J=1-0 amplification is larger than that of J=2-1 at both 
higher and lower densities, however the overall amplification factors are lower at higher densities. The weaker 
J=2-1 features may therefore  imply low densities of $\lse 4\times10^9$~H$_2\;$cm$^{-1}$ in the overlapping 
features, an argument used by \citet{Phillips:03} to explain similarly weak J=2-1 emission relative to J=1-0 in their 
observations of R Cassiopeiae.

The variation of amplification factor with density \citep{Doel:95} may also provide an explanation for the fact that 
most of the v=1 J=2-1 features do not have a v=1 J=1-0 counterpart. This trend is particularly visible in the west 
of the epoch~2 image in Figure~\ref{fig:I-overlay-epoch2}. If the masing gas in these regions falls in the intermediate density 
regime,  then the amplification of v=1 J=2-1 maser emission may be much larger than that of the v=1 J=1-0 maser 
emission in this region of the envelope, so that the v=1 J=1-0 maser emission is not significant in regions of v=1 J=2-1 
maser emission. This is not expected from the \citet{Gray:00} model, which predicts that the v=1 J=2-1 masers occur 
in a subset of the conditions where the v=1 J=1-0 masers occur.

Alternatively, the existence of v=1 J=2-1 masers without v=1 J=1-0 counterparts may be indicative of radiative
pumping of these masers. Under the predominantly radiative pumping model of \citet{Bujarrabal:94a}, v=1 J=2-1 
emission is stronger than the v=1 J=1-0 emission over most of the range of envelope conditions they investigated, 
with a greater difference at lower densities. However, as noted above, this relative ratio of J=2-1 and J=1-0 brightness 
is not observed across our data as a whole.

As mentioned above, J=2-1 emission is notably absent from region R1. Under the collisional pumping model of 
\citet{Doel:95}, the absence of v=1 J=2-1 emission in this region may imply lower density conditions in this region, 
leading to low amplification of the v=1 J=2-1 maser. However, if the close correspondence between the v=1 
\mbox{J=1-0} and v=1 \mbox{J=2-1} emission in this region is due to the effects of an H$_2$O line overlap, then 
the notable lack of v=1 J=2-1 emission could also place these masers in the higher density 
$\ga 10^{10}$~H$_2\;$cm$^{-1}$ region of the \citet{Soria-Ruiz:04} line-overlap model, where the v=1 and 
v=2 J=1-0 emission is coupled and the v=1 J=2-1 emission is absent.

Finally, the feature R4 (Figure~\ref{fig:I-overlay-epoch2} and Figure~\ref{fig:I-overlay-R3-R4}) in the epoch~2 maps is the 
only feature which is stronger and more spatially-extended in the J=2-1 line than the J=1-0 line. It extends over 
almost 10~km.s$^{-1}$ in velocity in the J=2-1 line, with an intense peak over the velocity range 29-32~km.s$^{-1}$ and an 
extended weaker tail up to $\sim36$~km.s$^{-1}$. The v=1 J=1-0 emission is located in the weaker tail of the v=1 J=2-1 emission,
and the feature is not detected in the v=2 \mbox{J=1-0} map.
R4 is also significantly linearly polarised in both of the SiO maser transitions observed: up to $\sim8\%$ in the v=1 \mbox{J=1-0} transition,
and $\sim19\%$ in the v=1 J=2-1 transition. The marked difference between the characteristics of this feature and the other overlapping 
features as discussed above may imply different physical conditions in this region of the gas, or a different excitation route for this feature.

\subsection{Circumstellar envelope morphology}
\label{subsec:Discussion_envelope}

The distribution of circumstellar SiO maser features is variable over time \citep{Diamond:03}, with individual maser 
features appearing, evolving, and disappearing, as envelope conditions and excitation conditions change. The 
distribution of maser features in single-epoch maps should therefore not be over-interpreted. Gaps in SiO maser 
rings, for example, can arise out of purely random maser spot configurations \citep{Gray:09}.

However, the trend of the v=1 and v=2 J=1-0 maser emission being concentrated predominantly to the east of 
VY~CMa (see Figures~\ref{fig:I-overlay-epoch1} and \ref{fig:I-overlay-epoch2}) persists over many VLBI observations of this source 
\citep{Miyoshi:03,Choi:08b,Zhang:12} and is seen also in the v=1 J=1-0 VLA map of \citet{Zhang:12}. The trend 
of more uniformly distributed v=1 J=2-1 emission, with some concentration in the southwest, is also observed in 
maps published earlier by \citet{Shibata:04}.

The persistence of these trends indicates that they may be connected to longer-term physical characteristics of the 
near-circumstellar envelope of VY CMa. 

Many authors have suggested that VY~CMa is surrounded by a circumstellar disk. Possible origins of a circumstellar 
disk include excess dust formation and mass loss above magnetic cool spots in equatorial stellar regions \citep{Soker:99a}, 
the effect of rotation \citep{Wittkowski:98}, the presence of a binary companion \citep{Cruzalebes:98}, or that a pre-main 
sequence disk has survived throughout the subsequent stellar evolution of VY~CMa \citep{Richards:98,Kastner:98}. 

The observed distribution of maser emission may be consistent with a disk geometry if the western and southwestern 
emission emanates from the polar region, where the flow of circumstellar material is faster and the gas is less dense. 
This would limit the line of sight velocity coherence, leading to the more compact maser features observed in this region.
The relatively large number of v=1 J=2-1 maser features in the polar region relative to the J=1-0 masers may be due to 
favourable envelope conditions for v=1 J=2-1 maser emission, as discussed in the previous section.

In this picture, the more extended north-eastern regions of epoch 2 J=1-0 emission which lie near the stellar velocity may 
fall in the equatorial plane and may be probing the disk around the star. 
This is illustrated in Figure~\ref{fig:D2-velocity}, which shows the the velocity structure of the epoch 2 v=1 J=2-1 image.
The extended emission in region R1, as well as the emission further from the star in the northeast, displays a velocity 
gradient away from the stellar velocity with distance from the star, consistent with a location in the forward lobe of a 
disk. The large extent of the emission in R1 in particular indicates 
favourable conditions for the J=1-0 masers in this region. This may imply greater velocity coherence in this region, or 
temperature and density conditions favourable to these masers. 

However, the morphology of the SiO maser emission can be explained in a variety of ways outside of a bipolar envelope 
model. \citet{Zhang:12} recently published a proper motion study of four epochs of v=1 J=1-0 SiO maser emission 
towards VY~CMa. The SiO masers display slow, quasi-spherical outflow, with no strong evidence for a bipolar outflow. 
They model six observed spoke-like SiO maser features using a ballistic-orbit model, which provides a good fit for 
most of the features. However, they note that the ballistic model assumes that the maser features are radially aligned, 
which is at odds with the expansion model they fit to the proper motion of the features. They suggest that a more 
realistic modelling of the complex near-circumstellar envelope may require acceleration driven by pulsation or giant 
convective cells and the use of a hydrostatic inner envelope \citep{Zhang:12}.

A number of recent observations of VY~CMa have shown that the circumstellar envelope is highly inhomogeneous 
and mass-loss from the star may occur in localised events \citep{Smith:01,Humphreys:05,Smith:09}. The variable 
characteristics of the VY~CMa SiO maser emission across the circumstellar envelope may be evidence of this localised 
enhanced mass-loss from the star. In this context, the sparser region of more compact masers in the southwest may 
be a region of enhanced mass loss, where the circumstellar material is more turbulent. This is consistent with the 
near-infrared Keck aperture-masking images of \citet{Monnier:99} which shows extended emission to the south and 
southwest of the star, possibly indicative of concentrated mass-loss in these directions.

\section{Conclusions}
\label{sec:Conclusions}

In this paper, we have presented VLBI images of the v=$\{$1,2$\}$ J=1-0 and v=1, J=2-1 SiO maser emission in total intensity, 
total fractional linear polarisation, and electric vector position angle toward the supergiant star VY CMa. The total 
intensity images in separate transitions were spatially aligned using a cross-correlation technique. The source VY~CMa 
was chosen because of its brightness across multiple SiO maser transitions. In the current paper we have examined 
the spatial overlap of individual features in the separate transitions, their relative brightness, and implications for 
maser pumping. In addition, we have analysed the observations as a whole to study the overall morphology of the 
near-circumstellar environment of VY CMa.

Our conclusions from the current work are: 

1) In epoch 2, in which the transition sensitivities are similar, the individual v=2, J=1-0 maser components are almost 
always found within a subset of the v=1, J=1-0 emission regions, and in the case of image region R1, with very close 
morphological similarity. This is consistent with the hydrodynamical model of \citet{Gray:00}. Although there are 
differences in overlap conditions across the source for these two transitions in detail, overall this finding in the current 
work is more supportive of collisional pumping rather than radiative pumping including an H$_2$O line overlap. 

2) The v=1 J=1-0 and v=1 J=2-1 epoch 2 maps display a greater level of coincident maser features than has 
previously been reported for these two transitions. However, in contradiction to the \citet{Humphreys:02b} model
prediction, the J=2-1 emission is weaker than the J=1-0 emission in most cases for the coincident features. 
This behaviour is consistent with the \citet{Doel:95} model, where the relative weakness of the v=1 J=2-1 emission 
implies a low density constraint.

3) The overall total intensity morphology of the SiO maser emission toward VY CMa exhibits long-lived large-scale 
features that persist over at least several years. They are very likely associated with the intrinsic physical structure 
of the near-circumstellar environment. Current data do not strongly constrain the bipolarity (or non-bipolarity) of 
the bulk kinematics, but there is evidence for local inhomogeneous or asymmetric mass-loss in this environment.

4) The current work demonstrates that, as well as providing constraints on maser pumping models, simultaneous 
observations of SiO maser emission in multiple transitions and full polarisation at the component level is a promising 
means to investigate both the theory of astrophysical masers as well as the environments of late-type evolved stars. 
Features such as R3 in epoch 2 demonstrate the power of this approach, and will be discussed in further detail in 
future papers.


\vskip 1cm

We thank the reviewer for numerous helpful recommendations which improved the clarity of the paper.
This material is based upon work supported by the National Science Foundation under Grant No. AST 0507473.


\bibliographystyle{mn2e}
\bibliography{biblist}

\appendix

\section{Map alignment}
\label{appendixA}

The SiO maser images from different transitions can be aligned by cross-correlating the two-dimensional 
peak intensity maps of the Stokes $I$ emission. The peak of the two-dimensional cross-correlation function gives the 
offset required to maximise the spatial correlation of the two 
maps. 
For increasing spatial overlap and structural similarity between the maps being aligned the central peak of the 
cross-correlation product will be increasingly compact and well-defined.

The measured offset is assumed to be the physical sky-plane offset between the maser emission from the two maps.
This assumption is justified if there are no systematic global linear offsets between the emission in the cross-correlated maps.
A global linear offset is not scientifically expected, as there is no large scale linear effect 
across the circumstellar envelope which could lead to an overall offset between emission from the different transitions.
From the circumstellar geometry of the maser emission, any possible offsets are more likely caused by 
effects with radial dependence. These might include maser pumping dependencies either on density or stellar 
radiation intensity.

When performing the cross-correlation alignment, it was found that extremely high intensity maser spots could bias the spatial
correlation; in this case the peak in the correlation product preferentially aligns the two highest spots in the two input 
images. In order to mitigate this, the peak intensity 
input maps were transformed to have unity intensity where the original intensity exceeded a specified 
cutoff level, and zero below this cutoff. 
The cutoff values were set to the contour levels shown in Figures~\ref{fig:7-2-3-P} to \ref{fig:A-P}:
ten times the maximum broadened noise variance $\sigma_I$ of the Stokes $I$ cubes for the epoch 1 data sets, and 
five times the maximum $\sigma_I$ for the epoch 2 data sets.
The maser maps were then cross-correlated pair-by-pair, and sub-pixel level offsets were determined through 
Gaussian fits to the dominant peak of each cross-correlation function. 


\begin{figure}
\includegraphics[width=41mm]{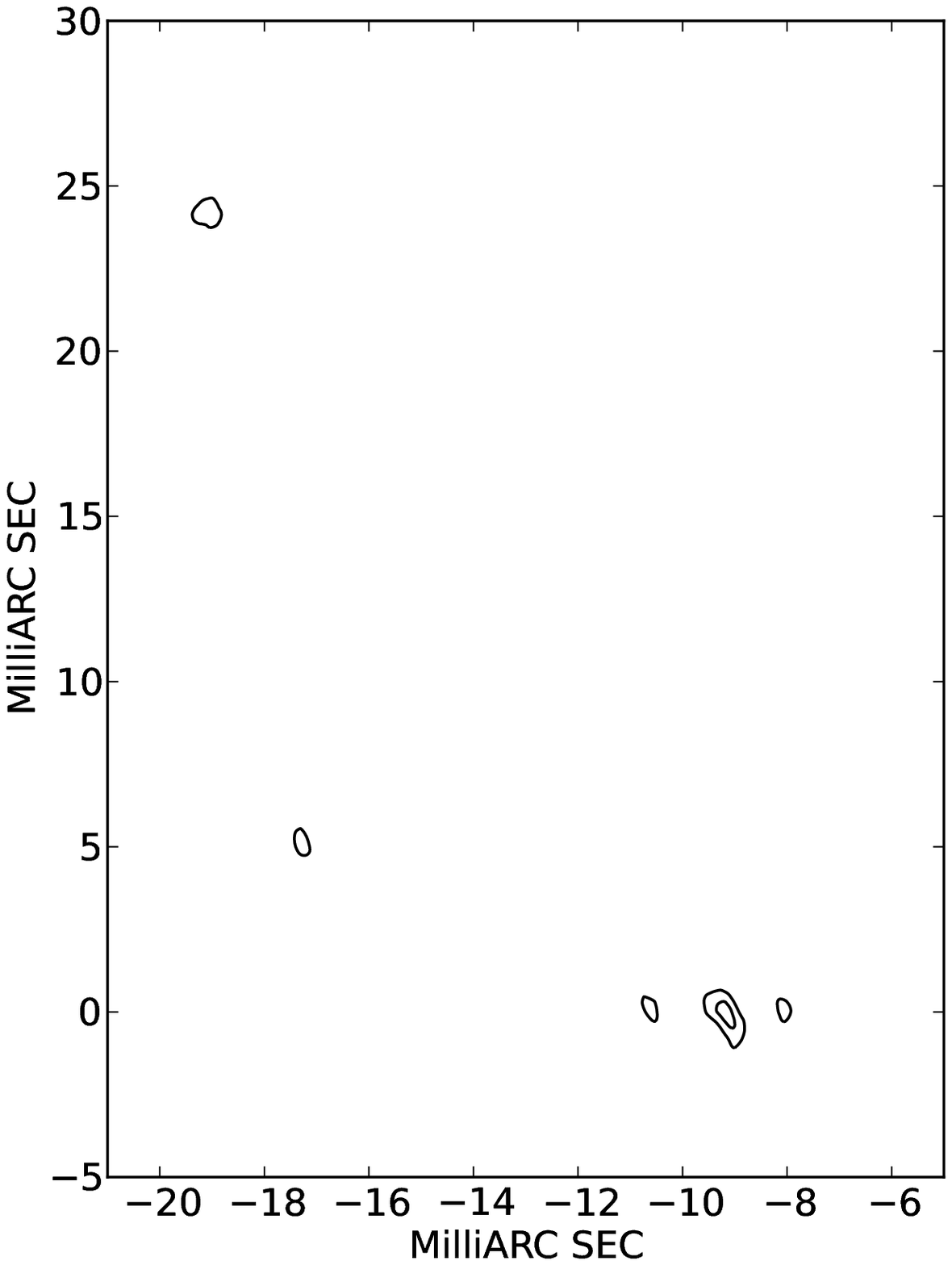}
\includegraphics[width=41mm]{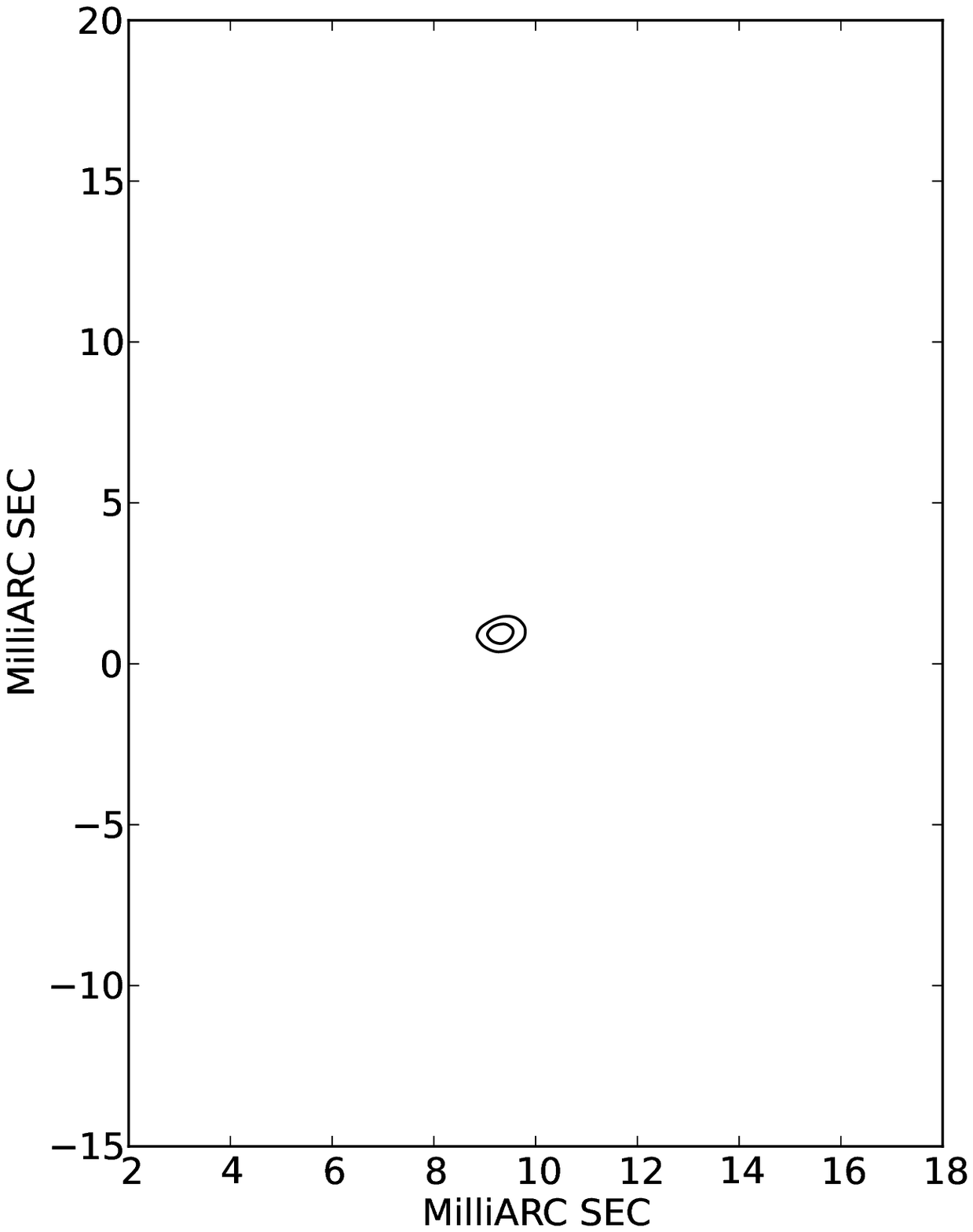}
\caption{Contour plots of the cross-correlation products, for the the epoch 1 v=1 J=1-0 and v=1 J=2-1 images
(left) and the \mbox{epoch 1} v=1 J=1-0 and v=2 J=1-0 images (right).
The contour levels are $70\%$ and $90\%$ of the peak value for each plot.}
\label{fig:correl}
\end{figure}


Examples of the cross-correlation functions are shown in Figure~\ref{fig:correl}. These plots show the regions
around the dominant peaks in the cross-correlation functions; the cross-correlation function was calculated
for the full extent of the images.
Figure~\ref{fig:correl} (left) shows the cross-correlation function for the v=1 J=1-0 and v=1 J=2-1 
images from epoch 1. This is an example of a case where the
correlation method performs poorly, as there is not sufficient overlapping emission in the maps to provide
a clear dominant compact peak in the cross-correlation product.
In this instance the results of the cross-correlation method were refined 
by manually testing the cross-correlation product peaks to determine
which peak is the correct offset between these images, considering the frequency distribution
of the emission as well as the maximum intensity Stokes $I$ image by visual inspection.  

The uncertainty in the offset is estimated as the ratio of the fitted Gaussian FWHM of the 
peak and the SNR of the source, following \citet{Phillips:03}. The minimum individual-channel SNR across all of the data 
cubes was $\lse 4$, and the overall maximum SNR was $\gg 100$. For the epoch~2 cubes the maximum SNR values were $\ga 400$. 
The uncertainty in the epoch 2 map alignment is therefore estimated to be $< 0.05$ mas. 
Assuming the correct peak has been selected in the epoch 1 alignment, the uncertainty 
is also estimated to be $< 0.05$ mas for epoch 1.


\end{document}